\documentclass[%
 reprint,
 amsmath,amssymb,
prd,
floatfix,
]{revtex4-2}
\usepackage{dcolumn,xcolor}

\usepackage{graphicx}
\usepackage{dcolumn}
\usepackage{bm}

\begin{document}


\title{Destabilization of Abelian-Higgs vortices at large Chern-Simons coupling}

\author{Emir S. Fadhilla$^{1}$}
\email{emir.syahreza@itb.ac.id}
\author{Mulyanto$^{1,3}$}
\email{muly031@brin.go.id}
\author{Hadi Susanto$^2$}%
\email{hadi.susanto@ku.ac.ae}
\author{Fiki T. Akbar$^{1}$}
\email{ftakbar@itb.ac.id}
\author{Bobby E. Gunara$^{1}$}
\email{bobby@itb.ac.id}

\affiliation{%
 $^1$Theoretical Physics Laboratory,
	Theoretical High Energy Physics Research Division,
	Faculty of Mathematics and Natural Sciences,
	Institut Teknologi Bandung
	Jl. Ganesha no. 10 Bandung, Indonesia, 40132
}
\affiliation{%
 $^2$Department of Mathematics, Khalifa University, PO Box 127788, Abu Dhabi, United Arab Emirates
}%

\affiliation{%
 $^3$Research Center for Quantum Physics, National Research and Innovation Agency (BRIN),
Kompleks PUSPIPTEK Serpong, Tangerang 15310, Indonesia
}%

\date{\today}

\begin{abstract}
We study the effect of the Chern--Simons (CS) coupling on the radial spectral stability of axially symmetric vortices in the $(2+1)$-dimensional Abelian--Higgs model. From the linear stability approach, we obtain the dispersion relations of gauge field perturbations, which signal the existence of a critical CS coupling \(k_{\rm c}\) that depends on Maxwell's coupling constant. For \(k<k_{\rm c}\), the continuous spectrum is purely imaginary and the modes remain gapped. At \(k=k_{\rm c}\), one of the modes becomes gapless, while for \(k>k_{\rm c}\) real eigenvalues appear, corresponding to exponentially growing radial modes. Direct time-evolution simulations are consistent with the spectral analysis, which shows instability for \(k>k_{\rm c}\). Below the threshold, the perturbed vortex relaxes toward its stationary profile, whereas above the threshold, growing fluctuations deform the core and propagate outward. These results show that the CS coupling controls radial spectral stability of Abelian-Higgs vortices, and destabilization occurs for sufficiently large CS coupling, relative to the Maxwell coupling.
\end{abstract}

\maketitle

\onecolumngrid
\section{Introduction}
Vortices in Abelian gauge--Higgs theories constitute a special class of finite-energy topological defects in relativistic field theory. Their origin can be traced to the flux-line solutions of the Ginzburg--Landau theory of superconductivity, where Abrikosov showed that the mixed phase of a type-II superconductor is described by quantized magnetic vortices \cite{Ginzburg:1950sr,ABRIKOSOV1957199}. The same mechanism was later reformulated in a Lorentz-invariant setting by Nielsen and Olesen, who constructed vortex-line solutions of the Abelian-Higgs model and interpreted them as stringlike defects of a spontaneously broken gauge theory \cite{Nielsen:1973cs}. In the relativistic Abelian-Higgs model, the finite-energy boundary conditions force the scalar field to approach the vacuum manifold at spatial infinity, while its phase winds around the vortex center. The corresponding map from the circle at spatial infinity into the vacuum manifold is classified by an integer winding number, and this integer is proportional to the magnetic flux. This topological property of Abelian-Higgs vortices is central in models of magnetic flux tubes, cosmic strings, and localized defects in gauge-scalar field theories \cite{Vilenkin:2000jqa,JaffeTaubes1980,Manton:2004tk}.

The critical Bogomolny–Prasad–Sommerfield (BPS) sector of the Abelian-Higgs model plays a special role in the theory of vortices. At this limit, the static energy functional saturates the BPS bound and admits a Bogomolny decomposition. This decomposition reduces the second-order Euler--Lagrange (EL) equations to a simpler first-order field equations known as Bogomol'nyi--Prasad--Sommerfield equations \cite{Bogomolny:1975de,prasad}. The construction of de Vega and Schaposnik gave an early explicit realization of Abelian-Higgs vortex solutions at the BPS limit \cite{deVegaSchaposnik1976}. A subsequent work then used an index-theoretic argument to count the zero modes of multivortex configurations, showing that a BPS vortex of winding number $n$ possesses $2n$ zero modes associated with the positions of the individual vortices \cite{Weinberg1979}. The existence and uniqueness of arbitrary $N$-vortex solutions of the first-order Ginzburg--Landau equations are then established in \cite{Taubes:1979tm}, thereby placing the multivortex model on rigorous mathematical ground. These developments led to the moduli-space description of low-energy vortex motion, in which BPS vortices behave as marginally bound objects whose slow dynamics can be approximated by geodesic motion on the vortex moduli space \cite{Ruback1988,Manton:2004tk}.

It is important to distinguish topological or BPS stability from dynamical spectral stability. In the former sense, a vortex is protected by its topological charge or by saturation of an energy bound. In the latter sense, one studies the spectrum of the linearized time-dependent equations about a given background and determines the presence or absence of exponentially growing perturbations. These two notions are related at the BPS limit but are not equivalent in general because the linear stability analysis can be performed for the second-order EL equations, which generally do not saturate the BPS bound. A spectral stability result also signals that the underlying time-dependent gauge-Higgs system has a well-defined evolution. The study of field dynamics beyond BPS limit is also backed by previous results proving the existence of finite energy solutions in finite time for several gauged-Higgs models, namely the Maxwell--Klein--Gordon model in $(2+1)$ dimensions \cite{Moncrief}; the Yang--Mills--Higgs model in four-dimensional Minkowski space \cite{EardleyMoncrief1982I,edrey};  Maxwell--Klein--Gordon model \cite{Klainerman}; and the MCSH model in $(2+1)$ dimensions \cite{chae}. These results provide the analytic background for treating perturbations of vortex solutions as part of a genuine dynamical field theory.

Planar gauge theories possess an additional structure that is absent in ordinary $(3+1)$-dimensional theories, namely the existence of the Chern--Simons (CS) interaction. For an Abelian gauge field in the effectively lower, $(2+1)$ dimensions, the CS density is proportional to $\epsilon^{\mu\nu\rho}A_\mu F_{\nu\rho}$. One of the interesting properties of this interaction is that it is first order in derivatives and metric independent, such that it essentially alters the dynamics of the gauge fields compared to the usual Maxwell's equations. It is shown that such a term generates a topological mass for the gauge field \cite{SCHONFELD1981157,Jackiw:1980kv}. The systematic formulation of topologically massive gauge theories was developed by Deser, Jackiw, and Templeton \cite{Deser:1981wh,Deser:1982vy}. When the CS term is coupled to a Higgs field, the Gauss constraint ties electric and magnetic sectors together, so that vortex configurations may carry electric charge as well as magnetic flux. This feature makes CS--Higgs theories a straightforward model for the study of charged planar solitons.

From the viewpoint of stability, the CS interaction has a particularly interesting status. Because the CS density is independent of the spacetime metric, its metric variation vanishes, and it gives no explicit contribution to the symmetric energy--momentum tensor \cite{Deser:1981wh}. Equivalently, it does not add a direct positive or negative term to the static energy functional in the way that the Maxwell term, the Higgs kinetic term, or the scalar potential do. Its influence on the energy of a solution can only arise indirectly, through the field profiles selected by the modified equations of motion. At the same time, variation with respect to the gauge potential produces an additional term proportional to the dual field strength, thereby modifying Gauss's law and the spatial gauge equations. Thus, the CS term is non-energetic at the level of the energy-momentum tensor but genuinely dynamical at the level of the EL equations, and such a situation has been discussed briefly in \cite{mulyanto2025global}. This separation motivates the present stability analysis, since a term that is invisible in the explicit energy density may nevertheless reorganize the perturbation spectrum of a localized Abelian-Higgs vortex.

Self-dual Abelian CS--Higgs vortices which saturate the BPS bound were initially introduced in \cite{Hong:1990yh,Jackiw:1990aw}. In contrast to Maxwell-Higgs vortices, these solutions arise in a theory where the CS term provides the gauge-field dynamics and where the electric charge is intrinsically linked to the magnetic flux. A further extension is obtained by combining Maxwell and CS terms in a single gauge-Higgs theory. The simultaneous presence of the Maxwell and CS terms generally obstructs the standard Bogomolny completion of the minimal gauge-Higgs model. However, it is shown that a self-dual Maxwell--Chern--Simons--Higgs (MCSH) theory can be obtained by introducing a real neutral scalar field \cite{Lee:1990eq}, and their related Bogomolny equations were later discussed in \cite{Lee:1991em}. In these models, the neutral scalar is not just an auxiliary modification but part of the mechanism that restores the self-dual structure and yields static charged vortices with controlled energy. The self-dual MCSH study has since been substantially generalized. In \cite{bazeia2012generalized}, a generalized self-dual MCSH framework with scalar-dependent couplings was constructed. Andrade, Bazeia, Marques, and Menezes developed nonminimal MCSH models supporting vortex configurations with modified core structure \cite{Andrade:2020nsd}. A subsequent work proposed a first-order framework for generalized MCS vortices without a neutral scalar field \cite{Andrade:2021uiu,AlonsoIzquierdo:2022zmw}. The BPS Lagrangian method introduced by Atmaja provides another systematic route for deriving first-order vortex equations \cite{atmaja2014bogomol,Atmaja:2015umo}, and it has been applied to generalized self-dual MCSH models and related classifications of BPS vortex sectors \cite{PrasetyaTama:2021elj}. These works clarify the static BPS structure of a wide family of planar Abelian-Higgs theories. They also provide natural comparison models for a dynamical stability analysis in which no BPS protection is assumed from the outset.

Specifically, in this work, we focus on the MCSH model in $(2+1)$-dimensional Minkowski spacetime described by
\begin{equation}\label{lagrangian}
{\cal L}
= -\frac{w}{4}F_{\mu\nu}F^{\mu\nu}
  -\frac{k}{4}\epsilon^{\mu\nu\rho}A_{\mu}F_{\nu\rho}
  + |D_{\mu}\phi|^2
  - m^2 V(|\phi|),
\end{equation}
where $w>0$ is the Maxwell coupling, $k$ is the CS coupling, $F_{\mu\nu}=\partial_\mu A_\nu-\partial_\nu A_\mu$, and $D_\mu\phi=\partial_\mu\phi+iA_\mu\phi$. The potential $V$ is left general in the analytic formulation, while explicit numerical calculations can be performed after choosing an explicit potential term. The model contains the Maxwell and CS terms together with a gauged complex Higgs field. The main approach is therefore spectral rather than Bogomolny-theoretic, namely the influence of the CS interaction on the linear perturbation problem for stationary vortex backgrounds.

The fluctuation spectra of Abelian-Higgs vortices provide the closest methodological precedent. Linearizing the field equations about a vortex background gives a fluctuation operator whose discrete and continuous spectra encode zero modes, internal modes, threshold behavior, and possible instabilities. The bound states and instabilities of global and local Abelian vortices have been discussed in \cite{GoodbandHindmarsh1995} using numerical matrix methods. Furthermore, the BPS vortex bound states were analyzed in \cite{AlonsoIzquierdoGarciaFuertesGuilarte2016PLB,AlonsoIzquierdoGarciaFuertesGuilarte2016JHEP}, which later gave a detailed description of zero modes and bound states by exploiting the hidden supersymmetric structure of the self-dual fluctuation operator. More recently, the normal-mode structure is also resolved for rotationally invariant $n$-vortices in the Maxwell-Higgs model with arbitrary scalar coupling, reducing the problem to radial spectral systems in angular momentum sectors \cite{AlonsoIzquierdoMiguelezCaballero2024}. Related studies of spectral flow, collective-coordinate dynamics, radiation from excited vortices, and scattering with excited normal modes demonstrate that internal vortex modes can have direct consequences for the nonlinear dynamics of vortices \cite{AlonsoIzquierdoGarciaFuertesMantonGuilarte2024,AlonsoIzquierdoMantonGuilarteWereszczynski2024,AlonsoIzquierdoBlancoPilladoEtAl2024,KruschReesWinyard2024}. This perspective is also consistent with the broader theory of soliton stability. Derrick's scaling argument gives a basic obstruction to finite-energy static scalar solitons in many relativistic models \cite{Derrick1964}. The Vakhitov--Kolokolov criterion and the Grillakis--Shatah--Strauss theory provide general spectral and variational frameworks for the stability and instability of solitary waves in Hamiltonian systems \cite{VakhitovKolokolov1973,GrillakisShatahStrauss1987,GrillakisShatahStrauss1990}. In vortex physics, semilocal and electroweak strings give important examples in which localized flux-tube solutions may exist while their stability depends on additional parameters or on the embedding of the defect in a larger field theory \cite{VachaspatiAchucarro1991,Hindmarsh1992,JamesPerivolaropoulosVachaspati1992}. These examples show that existence, topology, and static energy considerations do not by themselves determine the full dynamical response of a vortex background.

The main purpose of this work is to analyze the dynamical stability of MCSH vortices by means of linear perturbation theory. To formulate the corresponding radial spectral problem, we derive the coupled perturbed equations for fields around the stationary vortex background, and then linearize the equations by taking only up to the first-order perturbation terms. The asymptotic form of the perturbation equations is then used to identify the structure of the continuous spectrum, while a finite-radius discretization is used to approximate the spectrum of the radial linearized operator. We also integrate the time-dependent field equations numerically in representative parameter regimes in order to compare the evolution of the vortex profile under variations of parameters. This approach is designed to isolate the role of the CS coupling in the dynamical stability problem. The paper is organized as follows. In Sec. \ref{MCSHmodel}, we introduce the main field equations governing the dynamics in the MCSH model. In Sec. \ref{stationary}, we construct stationary vortex profiles and specify their boundary behavior. In Sec. \ref{stability}, we derive the linearized perturbation equations and identify the spectrum of the corresponding linearized dynamical system. In Sec. \ref{VortexTimeEvolution}, we present numerical time-evolution simulations of the vortex profile, and Section \ref{sec:conclusions} summarizes the analysis and discusses possible extensions. We adopt the Einstein summation convention with Greek indices $\mu, \nu, \rho, \sigma$ ranging over $0, 1, 2$ and Latin indices $i,j,k$ running over $1, 2$.

\section{Maxwell-Chern-Simons-Higgs model}\label{MCSHmodel}
 The field equations of motion derived from Lagrangian \eqref{lagrangian} are
 \begin{equation}\label{eom1}
     w\left(\partial_\sigma\partial^\sigma A^\mu-\partial^\mu\partial_\rho A^\rho\right)
     -k\epsilon^{\mu\rho\sigma}\partial_\rho A_\sigma-2\operatorname{Im}\left(\phi\overline{D^\mu \phi}\right)=0~,
\end{equation}
\begin{equation}
     \partial_\mu\partial^\mu\phi+2iA^\mu\partial_\mu \phi +i\phi\partial_\mu A^\mu-A^\mu A_\mu \phi+m^2{{\partial }_{\bar{\phi} }}V=0~.\label{eom2}
\end{equation}
In a more compact notation where we define ${{F}^{\nu }}\equiv\frac{1}{2}{{\epsilon }^{\nu \rho \sigma }}{{F}_{\rho \sigma }}$, the dual electromagnetic vector, the equations above reads
\begin{equation}
    w\partial_\nu F^{\nu\mu}=kF^\mu+2\operatorname{Im}\left(\phi\overline{D^\mu \phi}\right),~~~D_{\mu}D^{\mu}\phi+m^2{{\partial }_{\bar{\phi} }}V=0~.
\end{equation}

The $U(1)$ gauge transformations for both the scalar and the gauge fields are given by
\begin{eqnarray}
    \phi\to e^{i\chi}\phi \hspace{2em} A_\mu\to A_\mu-\partial_\mu\chi~.
\end{eqnarray}
We can replace $A_\mu$ with $ A_\mu-\partial_\mu\chi$ and derive the same $F_{\mu\nu}$. Correspondingly, we replace ${{D}_{\mu }}={{\partial }_{\mu }}+i{{A}_{\mu }}$ with $e^{i\chi}D_\mu e^{-i\chi}={{\partial }_{\mu }}+i\left( {{A}_{\mu }}-{{\partial }_{\mu }}\chi  \right)$. Therefore, if we replace $A_\mu$ with $ A_\mu-\partial_\mu\chi$ and $\phi$ with $e^{-i\chi}\phi$ we can obtain the same $D_\mu\phi$. The $U(1)$ gauge invariance can be observed directly in the field equations above. All these invariances can also be found at the Lagrangian level for Maxwell and scalar terms, but the \(U(1)\) gauge invariance of the CS term is not as straightforward due to the \(A_\mu\) factor, which does not enter the Lagrangian as a derivative. The action of the CS part of the model reads \(S_{CS}=-(k/4)\int\epsilon^{\mu\rho\sigma}A_\mu F_{\rho\sigma}~d^3x\) which transforms as \(S'_{CS}=-(k/4)\int\epsilon^{\mu\rho\sigma}(A_\mu-\partial_\mu\chi) F_{\rho\sigma}~d^3x=S_{CS}+(k/2)\int\epsilon^{\mu\rho\sigma}\partial_\mu\chi \partial_{\rho}A_{\sigma}~d^3x\). The second term is a boundary term after integration by parts by using \(\epsilon^{\mu\rho\sigma}\partial_\mu\chi \partial_{\rho}A_{\sigma}=\partial_\mu(\epsilon^{\mu\rho\sigma}\chi \partial_{\rho}A_{\sigma})\). The integral of this boundary term goes to zero if and only if the fields vanish at the asymptotic boundary. Thus, the \(U(1)\) gauge invariance is preserved by the CS term only if the fields vanish at the asymptotic boundary. This well-known property hints that models containing the CS term usually have localized classical solutions. Consequently, we are free to apply the gauge condition, and in this work we chose to apply the Lorenz gauge condition,
\begin{eqnarray}\label{LG}
    \partial^\mu A_\mu=0~.
\end{eqnarray}
When the Lorenz gauge condition is applied, the Maxwell-Chern- Simons-Higgs (MCSH) system transforms into a set of hyperbolic equations that exhibit a null-form structure.

Imposing the gauge condition \eqref{LG} to the field equations \eqref{eom1} and \eqref{eom2}, gives us the following dynamical system
\begin{eqnarray}\label{eomCG1}
    w\left(\partial_t^2-\nabla^2\right) {{A}^{0}}+k~\hat{\textbf{z}}\cdot\left(\nabla\times \textbf{A}\right)-2\operatorname{Im}\left[\phi\left(\partial_t-iA_0\right)\bar\phi\right]=0 ~,
\end{eqnarray}
\begin{eqnarray}\label{eomCG2}
    w\left(\partial_t^2-\nabla^2\right)\textbf{A}-k~\hat{\textbf{z}}\times\left(\partial_t\textbf{A}+\nabla A_0\right)+2\operatorname{Im}\left[\phi\left(\nabla+i\textbf{A}\right)\bar{\phi}\right]=0~,
\end{eqnarray}
\begin{eqnarray}\label{eomCG3}
   \left(\partial_t^2-\nabla^2\right)\phi+2iA_0\partial_t\phi+2i\textbf{A}\cdot\nabla\phi-\left((A_0)^2-\textbf{A}\cdot\textbf{A}\right)\phi+m^2\partial_{\bar{\phi}}V=0. 
\end{eqnarray}
Here, we also split the gauge potential $A_\mu$ into its time component $A_0$ and its spatial component $\textbf{A} = \left( A_i \right)_{i=1,2}$. For a function $\phi(t,x)$, we use $\nabla \phi = \left( \partial_i \phi \right)_{i=1,2}$ to denote its spatial gradient and $\partial \phi = \left( \partial_t\phi, \nabla \phi \right)$ to represent its full space-time gradient. The D’Alembertian operator, denoted by $\Box$, is defined as $\Box = \partial_t^2 - \Delta = \partial_t^2 - \partial_1^2 - \partial_2^2$.  

In the specific polar coordinate system, the wave equations read
\begin{align}\label{FUllDynamicsA0}
w(\partial_t^2-\Delta)A_0
+ k\left(\frac{1}{r}\partial_r(rA_\theta)-\frac{1}{r}\partial_\theta A_r\right)
- 2\operatorname{Im}\!\left[\phi(\partial_t-iA_0)\overline{\phi}\right] &= 0, \\[1ex]
w\left(\partial_t^2 A_r-\Delta A_r+\frac{A_r}{r^2}+\frac{2}{r^2}\partial_\theta A_\theta\right)
+ k\left(\partial_t A_\theta+\frac{1}{r}\partial_\theta A_0\right)
+ 2\operatorname{Im}\!\left[\phi(\partial_r+iA_r)\overline{\phi}\right] &= 0, \\[1ex]
w\left(\partial_t^2 A_\theta-\Delta A_\theta+\frac{A_\theta}{r^2}-\frac{2}{r^2}\partial_\theta A_r\right)
- k\left(\partial_t A_r+\partial_r A_0\right)
+ 2\operatorname{Im}\!\left[\phi\left(\frac{1}{r}\partial_\theta+iA_\theta\right)\overline{\phi}\right] &= 0, \\[1ex]
(\partial_t^2-\Delta)\phi
+ 2iA_0\partial_t\phi
+ 2i\left(A_r\partial_r\phi+\frac{A_\theta}{r}\partial_\theta\phi\right)
- \left((A_0)^2-A_r^2-A_\theta^2\right)\phi
+ m^2 \partial_{\overline{\phi}}V &= 0,\label{fullDynamicsPhi}
\end{align}
where the two dimensional Laplacian reads \(\Delta = \partial_r^2 + \frac{1}{r}\partial_r + \frac{1}{r^2}\partial_\theta^2\). To reduce the system into effectively \(1+1\) dimensional systems for \(A^\mu\), we later impose the condition for \(A_\mu\), \((A_0,A_r,A_\theta)=(A_0(t,r),A_r(t,r),A_\theta(t,r))\) on several parts of the analysis where only radial profiles are concerned. 

\section{Stationary solutions}\label{stationary}
Typically, stable solutions converge to stationary solutions at large \(t\). These stationary solutions are solved from eq. (\ref{FUllDynamicsA0}-\ref{fullDynamicsPhi}) by applying a certain ansatz. The ansatz for stationary solutions is constructed such that the time dependence in the equations is decoupled, resulting in an effefctively time-independent equations. Since we are interested in vortex solutions, the ansatz represents a localized profile that obeys axial symmetry.
To impose axial symmetry, we use the following well-known ansatz \cite{ABRIKOSOV1957199,Nielsen:1973cs}, which represents single vortex solutions pinned at the origin,
\begin{equation}\label{StationaryAnsatz}
	 A_r=0,\qquad A_\theta = -\frac{1}{r}\big(h(t,r)-n\big),\qquad A_0 = g(t,r),
\end{equation}
where \(n\in\mathbb{Z}\) is the integer winding number related to the quantization of magnetic flux on systems consisting of vortices.
This ansatz implies a divergenceless \(\textbf{A}\) which leads to \(\partial_tA_0=0\) under the Lorenz gauge. The resulting dynamical equations to be solved can then be recast into components such that the system becomes completely scalar equations.
The stationary radial solutions of \(\phi\) have a modulus that is independent of the angular coordinate, \(\partial_\theta|\phi|=0\). As such, we can express \(\phi\) as a radial real function multiplied by a phase with angular and temporal dependence as follows,
\begin{equation}\label{stationaryAnsatzScalar}
    \phi=f(r)e^{i\left(n\theta+\omega t\right)},
\end{equation}
An important condition that needs to be satisfied by \(f\) is that it must be zero at the origin to make \(\phi\) unique everywhere. Otherwise, the non-zero \(f(0)\) would make \(\phi\) become non-single-valued due to the angular phase. At large \(r\), \(f(r)\) usually goes to a constant value which minimizes the potential, \(V\). The stationary ansatz \eqref{StationaryAnsatz} comes with a set of boundary conditions. The boundary conditions are
\begin{equation}\label{ABoundCond}
    g(0)=g_0,~h(0)=n,~g(r\rightarrow\infty)=-\omega,~h(r\rightarrow\infty)=0,
\end{equation}
with \(g_0\in\mathbb{R}\) is a constant that usually becomes the shooting parameter to satisfy the conditions at large \(r\).
This set of boundary conditions is chosen so that the regularity at \(r=0\) is satisfied \cite{Atmaja:2018ddi,PrasetyaTama:2021elj}.
The ansatz \eqref{StationaryAnsatz} supplemented by expression \eqref{stationaryAnsatzScalar}, leads to the following stationary radial profile equations
\begin{eqnarray}\label{stationaryEqg}
    w\left(\partial_r^2g+\frac{\partial_rg}{r}\right)+\frac{k}{r}\partial_rh-2\left(\omega+g\right)f^2&=&0,\\
    w\left(\partial_r^2h-\frac{\partial_rh}{r}\right)-kr\partial_rg-2hf^2&=&0,\\
    \partial_r^2f+\frac{\partial_rf}{r}+\left[\left(\omega+g\right)^2-\frac{h^2}{r^2}\right]f-m^2\frac{dV}{df^2}f&=&0.\label{stationaryEqf}
\end{eqnarray}

When the CS coupling is turned off, \((k=0,w=1)\), we retrieve the Maxwell-Higgs (MH) case where the corresponding vortex solutions are well-known. 
The other interesting limit is the CS-Higgs (CSH) case where the Maxwell coupling is turned-off \((k=1,w=0)\).
In contrast to the standard MH case, the resulting system of equations for the CSH case is first order in \(h\) and \(g\). The solution of these equations are then used as the background solution to find the spectrum of linear perturbation in the stability analysis. The profile of the solution is similar to the usual Abrikosov vortex profile, that is the solution of MH case, with \(|\phi|\) equal to zero at \(r=0\) and monotonically increasing to \(|\phi|\rightarrow1\) as \(r\) increases. The introduction of the CS term only changes the steepness of the profile, where the vortex profile converges faster to its asymptotic value as \(k\) increases.

At large \(r\), the stationary equations (\ref{stationaryEqg}-\ref{stationaryEqf}) becomes simpler, namely
\begin{eqnarray}\label{AsymptoticSolg}
    w\partial_r^2g-2\left(\omega+g\right)f^2&\approx&0,\\
    w\partial_r^2h-kr\partial_rg-2hf^2&\approx&0,\\
    \partial_r^2f+\left(\omega+g\right)^2f-m^2\frac{dV}{df^2}f&\approx&0.\label{AsymptoticSolf}
\end{eqnarray}
For a specific potential \(V\) and boundary conditions, this system can be solved analytically. Since \(r\) takes values from zero to positive infinity, the solutions of (\ref{stationaryEqg}-\ref{stationaryEqf}) are close to the asymptotic solutions for most of the domain. As such, the asymptotic solutions are useful for the numerical approximation of the spectra of linear perturbations discussed in the next section for stability analysis.

\section{Stability of stationary solutions}\label{stability}
\begin{figure*}
    \centering
    \includegraphics[width=0.8\linewidth]{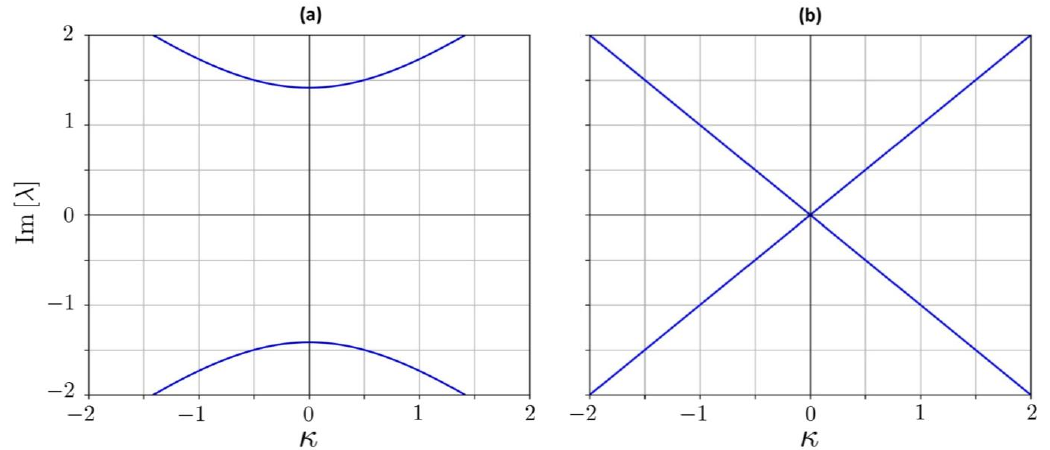}
    \caption{Dispersion relation of \(u\) \textbf{(a)} and \(v\) \textbf{(b)} for \(k=w=1\), \(m=1\) with \(\phi^4\) potential}
    \label{fig:DispersionScalar}
\end{figure*}

The solutions we found in section \ref{stationary} are stationary in the sense that if we start the system exactly with these stationary solutions, then the system will not evolve and the profile persists. If the system is stable, then any profile close enough to the stationary solutions will always evolve into these solutions given a sufficient interval of time, but if the system is unstable, the profile evolves away even for small deviations from the stationary solutions. In order to study the stability of the stationary solutions we found in the previous section, we introduce a perturbation around this solution in the following form
\begin{equation}
    \phi=\left(f(r)+\varepsilon u(t,r)+i\varepsilon v(t,r)\right)e^{i\left(n\theta+\omega t\right)}.
\end{equation}
Similarly, we also introduce time-dependent perturbation to \(A_0\) and \(A_i\)s, satisfying 
\begin{eqnarray}
    A_0&=&g(r)+\varepsilon a_0(t,r)~,\\ A_r&=&\varepsilon a_r (t,r)~,\\ A_\theta&=&-\frac{h(r)-n}{r}+\varepsilon a_\theta(t,r)~.
\end{eqnarray}
The constant \(\varepsilon\) is the perturbation parameter, which will be taken to be small for the linearization procedure. In this work, we focus on radial perturbations that depend only on \(t\) and \(r\), with no dependence on \(\theta\). This type of perturbation is imposed to maintain a similar structure to eq. \eqref{StationaryAnsatz}, since the general perturbation with angular dependence also requires a more general stationary ansatz.

Substitution of perturbed solutions above to the dynamical system (\ref{FUllDynamicsA0}-\ref{fullDynamicsPhi}) and isolating only up to first order equations in \(\epsilon\) gives us,
\begin{align}\label{linearizeda0}
w\left(\partial_t^2-\partial_r^2-\frac{1}{r}\partial_r\right)a_0
+k\,\frac{1}{r}\partial_r(ra_\theta)
+2f\,\partial_t v
+4(\omega+g) f\,u
+2f^2 a_0
&=0,
\\[0.8em]
w\left(\partial_t^2-\partial_r^2-\frac{1}{r}\partial_r+\frac{1}{r^2}\right)a_r
+k\,\partial_t a_\theta
+2\left(f'v-f\,\partial_r v+f^2 a_r\right)
&=0,
\\[0.8em]
w\left(\partial_t^2-\partial_r^2-\frac{1}{r}\partial_r+\frac{1}{r^2}\right)a_\theta
-k(\partial_t a_r+\partial_r a_0)
-\frac{4hf}{r}\,u
+2f^2 a_\theta
&=0,
\end{align}
\begin{align}
\partial_t^2 u-\partial_r^2 u-\frac{1}{r}\partial_r u
+\left(
\frac{h^2}{r^2}-(\omega+g)^2
+m^2\left[V'(f^2)+2f^2V''(f^2)\right]
\right)u&\nonumber
\\-2(\omega+g)\,\partial_t v
-2(\omega+g) f a_0
-\frac{2hf}{r}a_\theta
&=0,
\\[0.8em]
\partial_t^2 v-\partial_r^2 v-\frac{1}{r}\partial_r v
+\left(
\frac{h^2}{r^2}-(\omega+g)^2
+m^2V'(f^2)
\right)v
+2(\omega+g)\,\partial_t u
+2f' a_r
&=0,\label{linearizedv}
\end{align}
with the zeroth order perturbed equations being stationary equations (\ref{stationaryEqg}-\ref{stationaryEqf}) which also governs the profile of the background profile \((g,h,f)\). The resulting linearized equations constitute a coupled radial wave equations in two spatial dimensions that propagates in an axially symmetric medium with a radial profile depending on the vortex background. We study the spectrum of this system of coupled wave equations using asymptotic approximation and then proceed to the finite \(r\) case for the full spectrum through a numerical approach.

\subsection{Asymptotic behaviour}\label{largerspectrum}
\begin{figure*}[]
    \centering
    \includegraphics[width=0.8\linewidth]{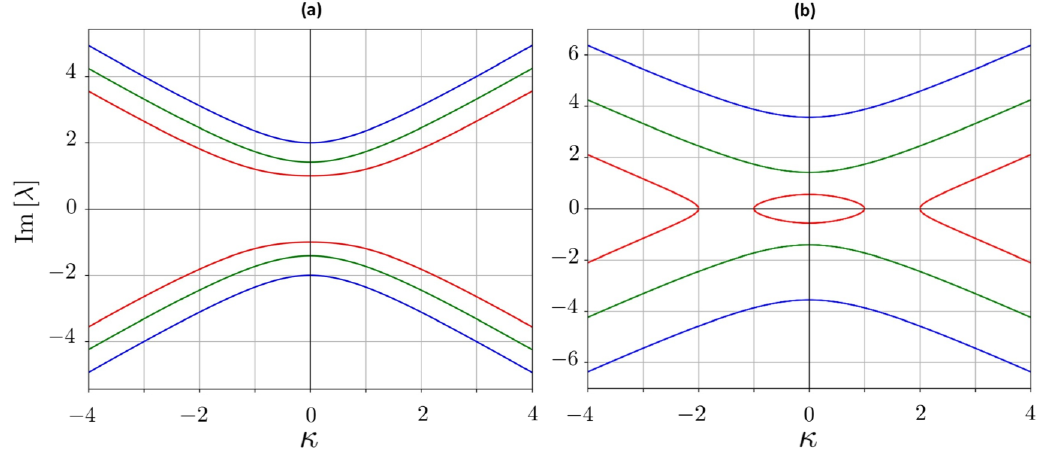}
    \caption{Dispersion relations of \(A^\mu\) with \(w=m=1\), \(k=1\) \textbf{(a)} and \(k=3\) \textbf{(b)} for \(\phi^4\) potential.}
    \label{fig:DispersionGauge}
\end{figure*}

At large \(r\), the wave equations satisfy the following approximations
\begin{align}
 \partial_t^2a_0
&=\partial_r^2a_0-\frac{k}{w}~\partial_ra_\theta
-\frac{2}{w}\left(f_\infty\,\partial_t v
+f_\infty^2a_0\right),
\\[0.8em]
\partial_t^2a_r
&=\partial_r^2a_r-\frac{k}{w}\,\partial_t a_\theta
+\frac{2}{w}\left(f_\infty\,\partial_r v-f^2_\infty a_r\right),
\\[0.8em]
\partial_t^2a_\theta
&=\partial_r^2a_\theta+\frac{k}{w}(\partial_t a_r+\partial_r a_0)
-\frac{2f^2_\infty}{w} a_\theta,\\
\partial_t^2 u
&=\partial_r^2 u
-2m^2f^2_\infty V''(f^2_\infty)
u,
\\[0.8em]
\partial_t^2 v
&=\partial_r^2 v,
\end{align}
where the asymptotic value of \(f(r)\), namely \(f_\infty\), satisfies \(V'(f_\infty^2
)=0\), implying that \(f_\infty\) is the critical point of the potential \(V\). Without any loss of generality, we can always rescale the scalar field such that \(f_\infty=1\). The asymptotic approximation of system (\ref{linearizeda0}-\ref{linearizedv}) yields a system of homogeneous coupled wave equations with some modification in the form of first derivatives. As such, we shall expect that the solutions of the asymptotic linearized equations are wave-like solutions that allows us to study the spectrum analytically through dispersion analysis. To proceed, we can derive the dispersion relation by assuming wave solutions with wave number, \(\kappa\), of the form \(q=(a_0,a_r,a_\theta,u,v)=e^{\lambda t+i\kappa r}(a_0^0,a_r^0,a_\theta^0,u^0,v^0)\), where \((a_0^0,a_r^0,a_\theta^0,u^0,v^0)\) is a constant vector, such that \(\partial_tq=\lambda q\).
This relation represents an eigenvalue problem which can be reduced into first-order autonomous equations by introducing \(p=(p_0,p_r,p_\theta,p_u,p_v)\equiv\partial_t q\).

In matrix form, this eigenvalue problem reads
\begin{equation}\label{asymptoticLinearized}
    \begin{bmatrix}
        \textbf{0}_{5\times5}&\textbf{1}_{5\times5}\\
        \mathcal{A}_q&\mathcal{A}_p
    \end{bmatrix}\begin{bmatrix}
        q\\p
    \end{bmatrix}=\lambda\begin{bmatrix}
        q\\p
    \end{bmatrix},
\end{equation}
where the matrices are explicitly
    \begin{equation}
    \mathcal{A}_q=\begin{bmatrix}
        -\kappa^2-\frac{2}{w}&0&-i\frac{\kappa k}{w}&0&0\\
        0&-\kappa^2-\frac{2}{w}&0&0&i\frac{2\kappa}{w}\\
        i\frac{\kappa k}{w}&0&-\kappa^2-\frac{2}{w}&0&0\\
        0&0&0&-\kappa^2-2m^2V''(1)&0\\
        0&0&0&0&-\kappa^2
    \end{bmatrix},\qquad\mathcal{A}_p=\begin{bmatrix}
        0&0&0&0&-\frac{2}{w}\\
        0&0&-\frac{k}{w}&0&0\\
        0&\frac{k}{w}&0&0&0\\
        0&0&0&0&0\\
        0&0&0&0&0
    \end{bmatrix}.
\end{equation}
Upon calculating the eigenvalues, \(\lambda\), there are four types of eigenvalues found, with one type containing two modes and the rest having one mode. These modes are, 
\begin{eqnarray}
    \lambda&=&\pm i\sqrt{\kappa^2+\frac{2}{w}}~,\\
    \lambda
&=&
\pm i\sqrt{
\kappa^2+\frac{2}{w}
+\frac{k^2}{2w^2}
+\frac{k}{2w^2}\sqrt{k^2+8w\bigl(1+\kappa^2 w\bigr)}
}~,\nonumber\\ 
\end{eqnarray}
\begin{eqnarray}
    \lambda&=&\pm i\sqrt{
\kappa^2+\frac{2}{w}
+\frac{k^2}{2w^2}
-\frac{k}{2w^2}\sqrt{k^2+8w\bigl(1+\kappa^2 w\bigr)}
}~,\nonumber\\ \label{Lambda3}\\
\lambda&=&\pm i\sqrt{\kappa^2+2m^2V''(1)}~,\\
\lambda&=&\pm i\kappa~.\label{lambda5}
\end{eqnarray}
An important observation is that the second and third modes above show up due to non-zero \(k\). The fourth mode corresponds to the dynamics of \(u\) while the fifth mode corresponds to the dynamics of \(v\), and we can observe that the dynamics of  \(u\) and \(v\) decouple from the system at large \(r\). As such, we can plot the eigenvalues of the scalar and gauge sector independently, while treating the terms containing \(v\) in the gauge sector as background. For the specific choice of  \(\phi^4\) potential, \(V=(1-|\phi|^2)^2/2\), the dispersion relation for \(u\) and \(v\) is plotted in Fig. \ref{fig:DispersionScalar}.
The dispersion relations for \(A^\mu\) are strictly gapped for small \(k\) but there exist finite \(k\) such that the gap of the third \(\lambda\) vanishes, as demonstrated in Fig. \ref{fig:DispersionGauge}. In the figure, the transition happens between the \(k=2\) (completely gapped) case and the \(k=3\) (ungapped) case. These dispersion relation is consistent under various gauge conditions. We demonstrate this property in Appendix \ref{GaugeInvariance} where we use gauge invariant fields to reproduce the asymptotic (large \(r\)) limit of eq. (\ref{linearizeda0}-\ref{linearizedv}).

The closing of the gap is actually a general property that is invariant under the choice of potential. We can show this, starting from the expression of the eigenvalues of the gauge field sector. The first and second eigenvalues are completely imaginary since there is no negative term inside the square root, but the third eigenvalue must satisfy 
\begin{equation}
    \kappa^2+\frac{2}{w}
+\frac{k^2}{2w^2}
-\frac{k}{2w^2}\sqrt{k^2+8w\bigl(1+\kappa^2 w\bigr)}\geq0,
\end{equation}
in order to be completely imaginary. Since \(k\) and \(w\) are positive, we can square both sides to have \(\kappa^4+\left[\left(4/w\right)-\left(k^2/w^2\right)\right]\kappa^2+(4/w^2)\geq0\), without any further constraint. We can see the LHS of this inequality as a quadratic form with respect to \(\kappa^2\), such that the inequality is satisfied if the LHS terms have at most one real root, i.e. \(\left[\left(4/w\right)-\left(k^2/w^2\right)\right]^2-(16/w^2)\leq0\). The last inequality implies that \((k^2/w)\geq0\), which is trivial because \(w\) is positive, and \((k^2/w)\leq8\) which sets the upper-bound for \(k\) in terms of \(w\). To conclude, the dispersion relation of the gauge field sector is gapped if
\begin{equation} \label{stableCondition}
    k\leq2\sqrt{2w}~~~~~(\text{linear stability condition}).
\end{equation}
If \(k\) is equal to or larger than this upper bound, then the gap closes. Beyond this bound, there exists \(\kappa\) such that \(\lambda\) is real. Furthermore, the real positive \(\lambda\) leads to instability, which hints that the stationary solutions with large enough CS coupling, \(k>2\sqrt{2w}\), are unstable against linear radial perturbation. 
\begin{figure*}
    \centering
    \includegraphics[width=0.85\linewidth]{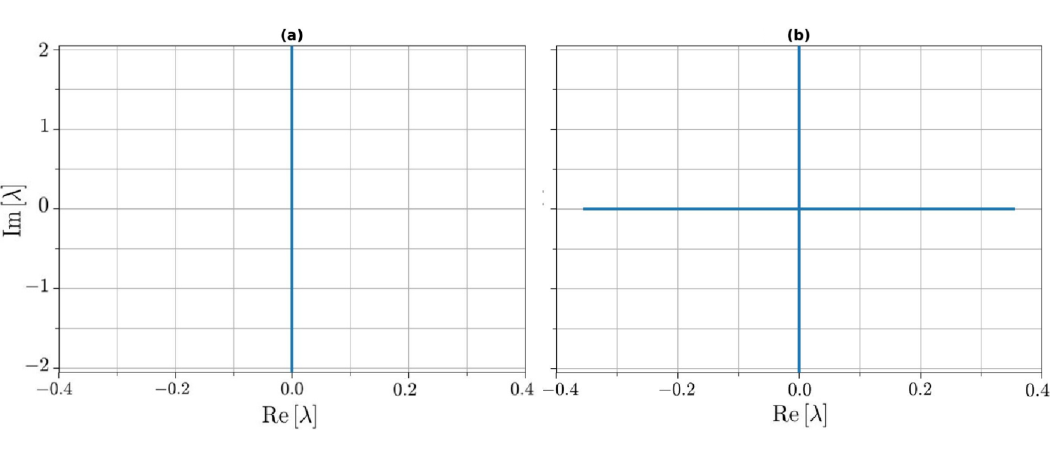}
    \caption{The continuous spectrum of eq. \eqref{asymptoticLinearized} for \textbf{(a)} stable \(k=w=1\) case and \textbf{(b)} unstable \(k=3,~w=1\) case.}
    \label{fig:ContinuousSpectrum}
\end{figure*}
\subsubsection{Linearly stable regime}
The linearly stable regime satisfies inequality \eqref{stableCondition}. In this regime, the eigenvalues of linear radial perturbations are all imaginary with no upper and lower bounds, i.e.
\begin{equation}
    \text{Im}[\lambda]\in\mathbb{R},~\text{Re}[\lambda]=0,
\end{equation}
as implied by the dispersion relation \eqref{lambda5}. This indicates the absence of exponentially growing modes in \(t\). An example of the spectrum of these stable modes is given in Fig. \ref{fig:ContinuousSpectrum}, where we plot the eigenvalues in the complex plane for the \(w=k=1\) case. 

As implied by the spectrum of \(\lambda\), in this linearly stable regime, the small perturbations are composed of harmonic forms, implying that any deviations that are close enough to the vortex profile will radiate away to large \(r\). This property can only be guaranteed at linear order of perturbations. However, it does not need a large perturbation to destabilize the vortex solutions. As implied by condition \eqref{stableCondition}, the transition to the unstable regime happens when we increase the coupling constant of the CS term, \(k\). The upper bound for stable values of \(k\) is strictly governed by Maxwell's term coupling, \(w\). 

Interestingly, this upper bound coincides with zero if Maxwell's term is turned off, namely \(w=0\). At this limit, \eqref{stableCondition} only allows \(k=0\) for stable vortices, which implies that the model becomes ungauged. At a glance, a straightforward physical conclusion is that all charged vortices are unstable at \(w=0\) limit, regardless of the value of \(k\). However, using the stable condition \eqref{stableCondition} to analyze vortex stability at \(w=0\) is actually inconsistent with the structure of the original dynamical system (\ref{FUllDynamicsA0}-\ref{fullDynamicsPhi}). When \(w\neq0\), the system is hyperbolic with both second derivatives in \(t\) and second derivatives in spatial coordinates, which allow us to use the dispersion relations to study the behaviour of the linearized (\ref{FUllDynamicsA0}-\ref{fullDynamicsPhi}) since the solutions are wave-like. On the other hand, at \(w=0\) limit, the system (\ref{FUllDynamicsA0}-\ref{fullDynamicsPhi}) becomes parabolic for the gauge fields with only first-order derivatives in \(t\). As such, we cannot apply the same treatment to the system. This failure of condition \eqref{stableCondition} at \(w=0\) limit is also hinted by the eigenvalues of the gauge sector, which are  undefined at this limit. Thus, we must rule out the CSH limit (\(w=0\)) for the condition \eqref{stableCondition} to work. The stability of specific \(n=\pm1\) vortices in the CSH model has actually been addressed in \cite{chen2009symmetric}, and the discussion of the stable spectra for this particular model will be discussed using different methods elsewhere. 
\subsubsection{Linearly unstable regime}
As shown above, the stationary solution becomes unstable when \(k>2\sqrt{2w}\). In this regime, there exist real eigenvalues solely contributed by \eqref{Lambda3}, while the rest of the eigenvalues are definitely imaginary for any \(k\) and \(w\). The instability of the stationary solutions comes from the positive real eigenvalues whose range of values can be directly deduced from the dispersion relation.

To use the dispersion relation to find the interval of positive real eigenvalues, we start from the condition for \eqref{Lambda3} to be completely real. This condition gives us 
\begin{equation}
    \kappa^2+\frac{2}{w}
+\frac{k^2}{2w^2}
\leq\frac{k}{2w^2}\sqrt{k^2+8w\bigl(1+\kappa^2 w\bigr)},
\end{equation}
which can be squared into \(\kappa^4+\kappa^2\left[\left(4/w\right)-\left(k^2/w^2\right)\right]+(4/w^2)\leq0\) because \(k\) and \(w\) are definite positive quantities. This inequality implies that \(\kappa^2\) satisfies \(\kappa_-^2\leq \kappa^2\leq\kappa_+^2\), where 
\begin{equation}
    \kappa_{\pm}^2\equiv\left(k^2/w^2\right)\left[\left(1\pm\sqrt{1-\left(8w/k^2\right)}\right)/2\right]-(2/w)
\end{equation} 
Notice that both the upper and lower bounds for \(\kappa^2\) are completely real for the unstable regime. As such, the intervals for \(\kappa\), where \(\lambda\) is real are
\begin{equation}\label{IntervalKappaRealLambda}
    \kappa\in\left[-\kappa_+,-\kappa_-\right]\cup\left[\kappa_-,\kappa_+\right].
\end{equation}
The values of \(\kappa\) which maximize \(\text{Re}[\lambda]\) must fall within these intervals, and the extreme values, namely \(\kappa=\kappa_\pm\), correspond to \(\text{Re}[\lambda]=0\). 
\begin{figure*}
    \includegraphics[width=0.85\linewidth]{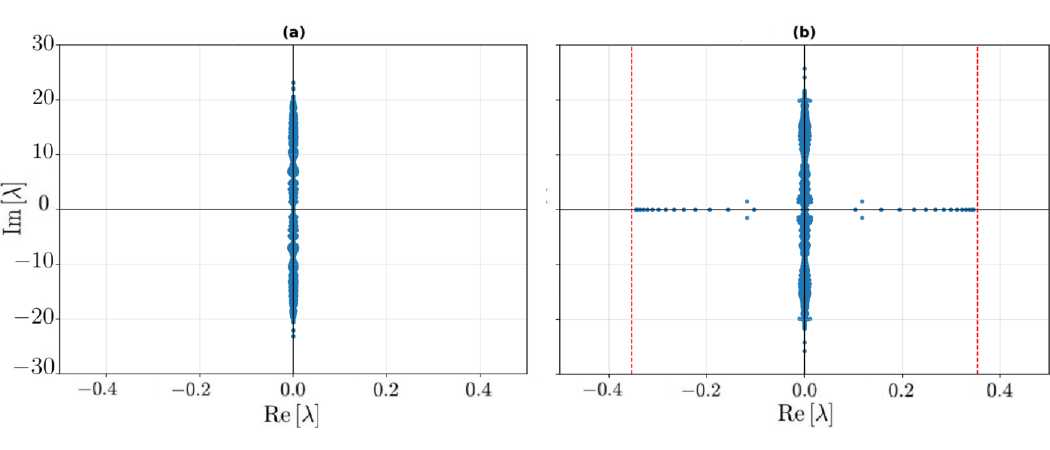}
    \caption{The real and imaginary parts of eigenvalues of linear perturbation at finite \(r\), for \textbf{(a)} stable \(k=w=1\) case, and \textbf{(b)} unstable \(k=3~,w=1\) case. The displayed spectrum assumes \(m=1\), \(n=1\), with \(\Delta r=0.1\) and large \(r\) boundary conditions are set at \(r_{\text{max}}=100\).}
    \label{fig:FiniteRSpectrum}
\end{figure*}

To find the upper and lower bounds of positive \(\text{Re}[\lambda]\), we can simply evaluate the derivative \(d\lambda^2/d(\kappa^2)|_{\kappa_m}\) to be zero which gives \(2k-\sqrt{k^2+8w(1+\kappa^2_mw)}=0\). This implies that \(\kappa^2_m=(3k^2-8w)/8w^2\) maximizes \(\lambda^2\). It is straightforward to observe that \(\kappa_m\) does fall into the intervals \eqref{IntervalKappaRealLambda}. Substituting the expression for \(\kappa_m\) to eq. \eqref{Lambda3} gives us the maximum and minimum of \(\text{Re}[\lambda]\), namely
\begin{equation}\label{EigValBound}
    \max\left[\text{Re}[\lambda]\right]=\sqrt{\frac{k^2-8w}{8w^2}},~~~\min\left[\text{Re}[\lambda]\right]=-\sqrt{\frac{k^2-8w}{8w^2}}.
\end{equation}
The \(\lambda\) above is a real definite quantity for the unstable regime, \(k> 2\sqrt{2w}\). As a conclusion, the real positive eigenvalues corresponding to the unstable modes belong to the interval \(0\leq\text{Re}[\lambda]\leq\sqrt{(k^2-8w)/(8w^2)}\). We are going to use this inequality in the convergence test for the numerical approach we use for the finite \(r\) case in the following section (See also Appendix section \ref{convergence}). An example of the spectrum of these unstable modes is given in Fig. \ref{fig:ContinuousSpectrum} where we plot the eigenvalues in the complex plane for \(w=1,~k=3\) case.

\subsection{Finite \(r\) behaviour}\label{finiterspectrum}
The first-order autonomous equations found in the asymptotic approximation are useful to simplify the eigenvalue problem. Similarly, for this more general case with finite \(r\),
the second-order eq. (\ref{linearizeda0}-\ref{linearizedv}) can be recast into simpler first-order equations using vector \(p\). In terms of vector \((q,p)\), the operator \(\partial_t\) becomes a \(10\times10\) matrix containing operator \(\partial_r\) which satisfies the first-order equations above. In matrix notation, this reads
\begin{equation}\label{finitersystem}
    \partial_t\begin{bmatrix}
        q\\p
    \end{bmatrix}=\begin{bmatrix}
        \textbf{0}_{5\times5}&\textbf{1}_{5\times5}\\
        \mathcal{B}_q&\mathcal{B}_p
    \end{bmatrix}\begin{bmatrix}
        q\\p
    \end{bmatrix}~,
\end{equation}
where the matrices \(\mathcal{B}_q\) and \(\mathcal{B}_p\) are explicitly given by,
    \begin{equation}
    \mathcal{B}_q=\begin{bmatrix}
        \mathcal{D}_r-\frac{2f^2}{w}&0&-\frac{k}{w}\left(\frac{1}{r}+\partial_r\right)&-\frac{4(\omega+g)}{w}f&0\\
        0&\mathcal{D}_r-\frac{1}{r^2}-\frac{2f^2}{w}&0&0&-\frac{2}{w}\left(f'-f\partial_r\right)\\
        \frac{k}{w}\partial_r&0&\mathcal{D}_r-\frac{1}{r^2}-\frac{2f^2}{w}&\frac{4hf}{wr}&0\\
        2(\omega+g)f&0&\frac{2hf}{r}&\mathcal{D}_r+\Gamma-m^2\left(V'+2f^2V''\right)&0\\
        0&-2f'&0&0&\mathcal{D}_r+\Gamma-m^2V'
    \end{bmatrix}~,
\end{equation}
\begin{equation}
    \mathcal{B}_p=\begin{bmatrix}
        0&0&0&0&-\frac{2f}{w}\\
        0&0&-\frac{k}{w}&0&0\\
        0&\frac{k}{w}&0&0&0\\
        0&0&0&0&2(\omega+g)\\
        0&0&0&-2(\omega+g)&0
    \end{bmatrix}~.
\end{equation}
In the components of the matrices above, we have introduced the operator \(\mathcal{D}_r\equiv\partial_r^2+(1/r)\partial_r\) and a background function \(\Gamma=\Gamma(r)\equiv (\omega+g)^2- (h^2/r^2)\) as shorthand notations. To find the spectrum of the system above, again we impose \((q,p)\propto e^{\lambda t}\) such that the time derivative operator satisfies the eigenvalue equation \(\partial_t(q,p)=\lambda(q,p)\). The matrices \(\mathcal{B}_q\) and \(\mathcal{B}_p\) contain derivatives with respect to \(r\) in their components, which can be recast into block matrices for numerical approximation, using finite difference with a certain slice \(\Delta r\). For the first derivative of any functions, \(\mathcal{F}(t,r)\), with respect to \(r\) we use the standard forward finite difference, \(\Delta r\partial_r\mathcal{F}\simeq \mathcal{F}(t,r+\Delta r)-\mathcal{F}(t,r)\). The second-order derivatives with respect to \(r\), which arise from the Laplacian, always show up in the form \(\partial_r^2\mathcal{F}+\partial_r\mathcal{F}/r\) that is equivalent to \(\partial_r(r\partial_r\mathcal{F})/r\). Applying finite difference to this operator gives us \(\Delta r^2\partial_r(r\partial_r\mathcal{F}(t,r ))/r\simeq\mathcal{F}(t,r+\Delta r)-\mathcal{F}(t,r )-\left(r-\Delta r\right)\left(\mathcal{F}(t,r )-\mathcal{F}(t,r-\Delta r)\right)/r\).

The system (\ref{FUllDynamicsA0}-\ref{fullDynamicsPhi}) and the dynamical system of its linear perturbation contain terms that are inversely proportional to \(r\), which cannot be solved directly from the origin, \(r=0\). Furthermore, the asymptotic boundary conditions at \(r\rightarrow\infty\) can only be numerically evaluated at finite large \(r\). As such, we introduce small radius \(r_{\text{min}}=\Delta r/2\) and large maximum radius, \(r_{\text{max}}\) such that the radial slice satisfy \(\Delta r=r_{\text{max}}/(N_{\text{grid}}-1)\) where \(N_{\text{grid}}\) is the number of points in the grid. The equations are then solved in the interval from \(r_{\text{min}}\) to \(r_{\text{max}}\). The \(r_{\text{min}}\) is chosen to be half \(\Delta r\) to handle the radial part of the Laplacian at the origin. The radial part of the Laplacian contains \(\left(r-\Delta r\right)\left(\mathcal{F}(t,r )-\mathcal{F}(t,r-\Delta r)\right)\) term, containing \(r-\Delta r\) which can be negative in the usual interval of \(r\). This term at \(r_{\text{min}}\) evaluates to \(\left(-\Delta r/2\right)\left(\mathcal{F}(t,r_{\text{min}} )-\mathcal{F}(t,r_{\text{min}}-\Delta r)\right)\). Since \(r\geq0\), then in this scheme we shall take \(r_{\text{min}}-\Delta r\) to be zero. As a result, the second term in the radial part of the Laplacian is zero at \(r_{\text{min}}\) and it evaluates to \(\Delta r^2\partial_r(r\partial_r\mathcal{F}(t,r ))/r\simeq\mathcal{F}(t,r_{\text{min}}+\Delta r)-\mathcal{F}(t,r_{\text{min}} )\).

It is found that the spectrum of eigenvalues of the matrix \eqref{finitersystem} resembles the continuous spectrum found from asymptotic approximation. For the linearly stable case where \(k\leq2\sqrt{2w}\), most of the eigenvalues have a large imaginary part compared to their real part, which places them close to the imaginary axes. These small real parts can be considered as small errors that arise from the discretization of the operator \(\partial_r\). For the linearly unstable case, \(k>2\sqrt{2w}\), we found that the spectrum also resembles its continuous counterpart. The distribution of harmonic modes on the imaginary axes is similar to the stable case, where most of the real parts are much smaller than their imaginary part. The unstable modes with large real parts are placed exactly on the real axes, which implies that these modes do not have harmonic parts. The eigenvalues of these unstable modes are bounded by the same bound \eqref{EigValBound} that is deduced from the continuous spectrum. In conclusion, the spectrum for finite \(r\) confirms the continuous spectrum we found from the asymptotic approximation. 

\begin{figure*}
    \centering
    \includegraphics[width=0.8\linewidth]{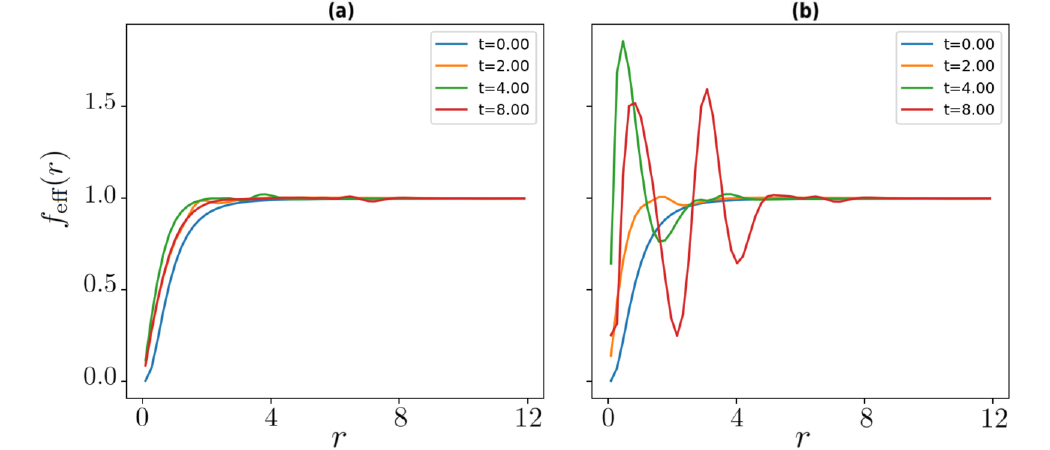}
    \caption{The evolution of the effective profile of the scalar field plotted against the radial coordinate for \textbf{(a)} \(k=1,w=1\) case and \textbf{(b)} \(k=3,w=1\) case. Both profiles are simulated with the same \(t_{\text{max}}\) using the characteristic time of \(k=3,w=1\) case.}
    \label{fig:profileScalar}
\end{figure*}
The spectrum we found numerically for the case of \(w=1\) and \(k=1,3\) is plotted in Fig. \ref{fig:FiniteRSpectrum}, with \(\Delta r=0.1\) and the potential is chosen to be the \(\phi^4\) potential. We can see that the small deviation from the continuous spectrum happens for eigenvalues with a large imaginary part. This small deviation most likely came from the sensitivity of the derivative operators inside the matrix \eqref{finitersystem} under coordinate discretization for the numerical procedure,  which is supported by the distribution of pseudo-spectra explained in detail in the Appendix section \ref{convergence}. Eq. \eqref{finitersystem} is most insensitive near the real axes. Thus, the instability of stationary solutions at large \(k\) comes from these unstable modes with zero harmonic parts, while the small real parts near the imaginary axes can be neglected. 

\section{Time evolution of vortex profile}\label{VortexTimeEvolution}
In the previous section, we discussed the conditions in which stationary vortex solutions become unstable. Thus, it is of interest to study how the profile of the vortex evolves over time for both stable and unstable regimes. To deal with the complex nature of the scalar field, we decompose the scalar field into its modulus and a phase factor that depends only on \(t\) and \(\theta\), as follows 
\begin{equation}
    \phi\equiv F(t,r,\theta)e^{i\Gamma(t,\theta)},
\end{equation}
which satisfies \(F=\sqrt{\phi_\text{re}^2+\phi_\text{im}^2}\) and \(\Gamma=\tan^{-1}\left(\phi_\text{im}/\phi_\text{re}\right)\) if \(\phi\) is written as \(\phi=\phi_\text{re}+i\phi_\text{im}\). During the numerical procedure, the scalar field is calculated in both components \(\phi_\text{re}\) and \(\phi_\text{im}\), and then recast in \(F\) and \(\Gamma\) for evaluation.
 For the time evolution solver in this section, we adopt the same definitions for \(r_{\text{min}}\) and \(r_{\text{max}}\) from Section 4.2. The details on the numerical procedure for the time evolutions are provided in Appendix section \ref{timeIntegrateMethod}.

 It is claimed in Section 4.2 that the spectrum of \eqref{finitersystem} resembles the continuous spectrum and the small real parts of the eigenvalues near the imaginary axes are negligible, resulting in linearly stable vortices with \(k<2\sqrt{2w}\). We demonstrate this stability condition by simulating the time evolution of the vortex profile up to a certain \(t_{\text{max}}>t_{\text{char}}\). The characteristic time, \(t_{\text{char}}\), here can be approximated from the upper bound of \(\text{Re}[\lambda]\) in \eqref{EigValBound}, such that
 \begin{equation}
     t_{\text{char}}=\begin{cases}
         0&,~k\leq2\sqrt{2w}\\
         \sqrt{\frac{8w^2}{k^2-8w}}&,k>2\sqrt{2w}
     \end{cases}\qquad.
 \end{equation}
This choice of \(t_{\text{max}}\) came from the fact that, if the vortex is stable, the profile always evolves to its stationary vortex profile at large \(t\), regardless of the \(t_{\text{max}}\), which implies that there is no characteristic time to observe the effect of small fluctuations on the profile. On the other hand, if the vortex is unstable, the fluctuation grows, with the fastest rate being the maximum of \(\lambda\). As such, when \(t<t_{\text{char}}\), the deviation from the stationary profile is difficult to observe, while when \(t>t_{\text{char}}\), the fluctuations become large enough to be observed. In Fig. \ref{fig:profileScalar} we plot the root-mean-square (RMS) profile of the scalar fields for several time stamps in the case of \(w=1\) with both stable \(k=1\) and unstable \(k=3\). The angular RMS scalar profile \(f_{\text{eff}}\) is defined through the modulus \(F\) by average over the whole circle at a certain \(r\), explicitly \(f_{\text{eff}}=\left(\int_0^{2\pi}F^2(t,r,\theta)d\theta/(2\pi)\right)^{1/2}\). In this figure, we can observe that for the stable case, the profile eventually gets close to the stationary profile as time increases. In contrast to the stable case, for the unstable case, the profile fluctuates with higher deviation as time increases. This dynamical behaviour of the vortex profile confirms our prediction from the spectral analysis, where unstable modes appear when \(k\) is sufficiently large, namely \(k>2\sqrt{2w}\).

\begin{figure*}
    \centering
    \includegraphics[width=0.95\linewidth]{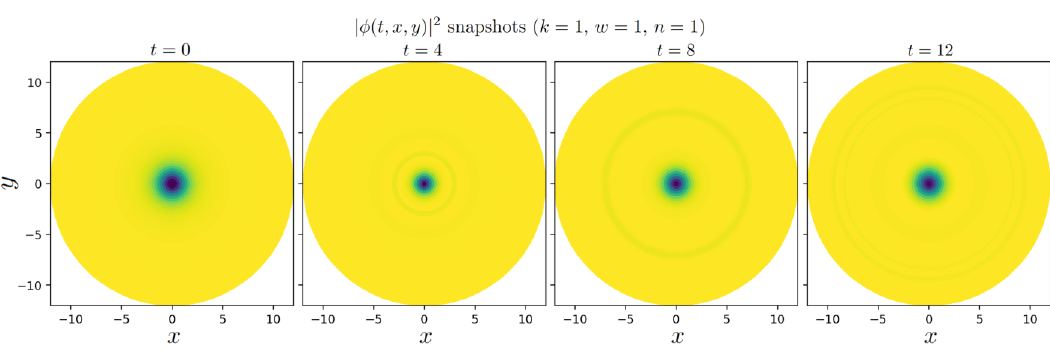}
    \includegraphics[width=0.95\linewidth]{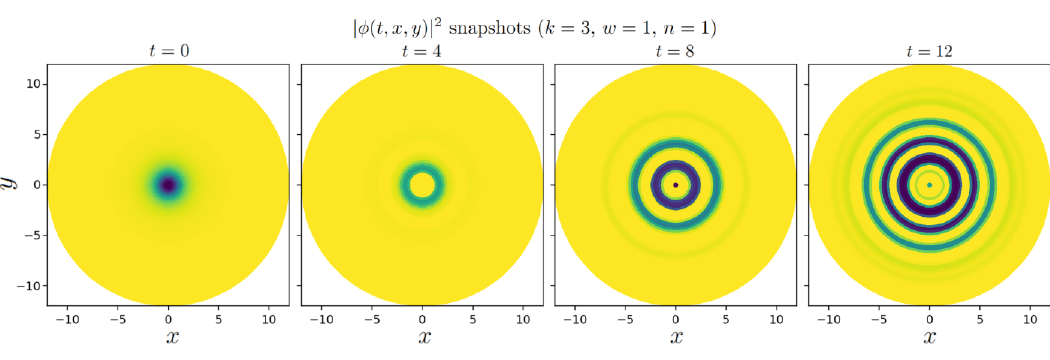}
    \caption{Profile of \(|\phi|^2\) with \(n=1\) for both stable (\(k=1,w=1\)) and unstable case (\(k=3,w=1\)).}
    \label{fig:2DvortexProfile}
\end{figure*}

To better observe how the fluctuations behave over time evolution of the vortex, we also plot the profile of \(F^2\) over the two-dimensional plane for several time stamps in Fig. \ref{fig:2DvortexProfile}. As seen in the figure, the small fluctuations decay over time for \(k<2\sqrt{2w}\), leaving the stationary profile intact. On the other hand, for larger \(k\) with \(k>2\sqrt{2w}\), the vortex core structure at the origin is suppressed by the growing fluctuations . 

The destabilization of local solutions \(k\) can also alter the topological degree of the scalar field. This is usually indicated by changes in the topological winding number of the vortex that is defined by
\begin{eqnarray}
    Q(t)&=&\frac{i}{2\pi}\int\epsilon^{0jk}\partial_j\phi\partial_k\bar{\phi} ~d^2x\nonumber\\&=&\frac{1}{\pi}\int_0^\infty\int_0^{2\pi}~\text{Im}\left[\partial_r\bar{\phi}\partial_\theta\phi\right] ~d\theta dr.
\end{eqnarray}
However, our simulation shows that the winding number, \(Q(t)\), for both \(k=1,w=1\) and \(k=3,w=1\) cases does not show significant changes over time, i.e. \(Q(t)\approx n\) for any \(t\). As such, the destabilization from large \(k\) coupling does not generate a topological transition in our simulation. We would like to note that the system considered in this simulation is a single vortex system where the vortex is pinned at the origin. Thus, the possibility of a topological transition in the destabilization of a more general system with multiple vortices is still open, but the proof for this behaviour requires a more complex setup and rigorous analysis beyond the scope of this work.

\section{Conclusions and outlook}
\label{sec:conclusions}

We have analyzed the radial spectral stability of axially symmetric vortices in the $(2+1)$-dimensional MCSH model with \(w>0\). We linearized the time-dependent field equations by introducing perturbations around stationary vortex profiles as backgrounds, and formulated the corresponding radial spectral problem from these linearized equations. The asymptotic perturbation equations show that the gauge-sector spectrum is purely imaginary for \( k < 2\sqrt{2w}\). At the critical value \(k=2\sqrt{2w}\), one gauge branch becomes gapless, and for \(k>2\sqrt{2w}\) this branch contains real eigenvalues. The associated positive real eigenvalues represent exponentially growing radial modes. Thus, within the radial linearized problem, a sufficiently large CS coupling destabilizes the vortex, even though the CS term generally does not contribute directly to the static energy density.

The finite-radius numerical spectrum supports this asymptotic criterion. For the \(\phi^4\) potential, the discretized radial operator has a spectrum that agrees with the continuous spectrum prediction. In the linearly stable regime, the physical spectrum lies on the imaginary axis up to discretization effects, whereas above the threshold, robust real eigenvalues appear on the real axis. Time-dependent simulations provide an independent check of this picture. For \(k<2\sqrt{2w}\), the perturbed profile relaxes toward the stationary vortex. For \(k>2\sqrt{2w}\), growing fluctuations deform the vortex core and propagate outward. In the single-vortex, axially symmetric simulations considered here, the winding number remains conserved within numerical accuracy. The instability observed in this setting is therefore a dynamical instability of the vortex profile rather than a topological transition. Thus, the time-dependent simulations show that, in the regime where \(k\) is less than or equal to the critical value, the vortex profile is stable under small linear perturbations. We note that the absence of exponentially growing modes in this regime does not generally imply full nonlinear stability, and we address this generalized problem elsewhere.

Several extensions are natural. The spectral analysis can be generalized to perturbations with angular dependence, for example, by adding angular phase on \((u,v)\) and deforming both \(g\) and \(h\) using harmonic functions in \(\theta\). This would determine whether non-radial modes modify the stability criteria and clarify the roles of the winding number \(n\) and the frequency parameter \(\omega\). A gauge-invariant formulation of the perturbation problem, together with tests for other Higgs potentials, would help determine which aspects of the stability criterion are universal. Furthermore, vortices with large \(n\) and multivortex configurations should be evolved without pinning the core at the origin, since unstable non-radial modes may lead to vortex splitting while preserving the total winding. Such properties have not been observed in our simulation since our method using polar coordinate pins the vortex at \(r=0\). 

An important limit that has not been covered by criterion \eqref{stableCondition} is the pure CSH limit. The linear stability at this limit, where \(w=0\), requires a separate analysis because the gauge-field equations no longer have the hyperbolic structure used in the present dispersion calculation. This problem will be addressed in future work.

\vfill

\begin{acknowledgments}
This research is funded by the Indonesian Endowment Fund for Education (LPDP) on behalf of the Indonesian Ministry of Higher Education, Science and Technology and managed under the EQUITY Program (Contract No. 4298/B3/DT.03.08/2025).
\end{acknowledgments}
\appendix

\section{Gauge invariance of the dispersion relation at large \(r\)}\label{GaugeInvariance}
Since the modulus, \(|\phi|\), does not transform under \(U(1)\) transformation, let us recast the Higgs field \(\phi\) as \(\phi=|\phi| e^{i \arg \left[\phi\right]}\) where \(\arg\left[\phi\right]\equiv \tan^{-1}\left(\text{Im}(\phi)/\text{Re}(\phi)\right)\). As implied by this decomposition, the only component that transforms under \(U(1)\) in \(\phi\) is the phase \(\arg [\phi]\), and the strategy to construct a gauge-invariant field is to absorb this transformation into a new field, \(C_\mu\equiv A_\mu+\partial_\mu\arg[\phi]\). Under \(U(1)\), the gauge field transforms as \(A_\mu\rightarrow A_\mu-\partial_\mu\chi\) while the phase transforms as \(\arg [\phi]\rightarrow\arg[\phi]+\chi\), hence the new field \(C_\mu\) transforms as 
\begin{equation}
    C_\mu\rightarrow \left(A_\mu-\partial_\mu\chi\right)+\partial_\mu\left(\arg [\phi]+\chi\right)=C_\mu.
\end{equation}
In other words, the new field \(C_\mu\) is not affected by the gauge transformation, similar to \(|\phi|\), implying that both fields are gauge-invariant.
Substituting the decompositions \(\phi=|\phi| e^{i\arg [\phi]}\) and \(A_\mu=C_\mu-\partial_\mu\arg [\phi]\) into \eqref{eom1} and \eqref{eom2} gives
\begin{eqnarray}
    w\left(\partial_\sigma\partial^\sigma C^\mu-\partial^\mu\partial_\rho C^\rho\right)
     -k\epsilon^{\mu\rho\sigma}\partial_\rho C_\sigma+2|\phi|^2C^\mu&=&0~,\nonumber\\
     \partial_\mu\partial^\mu|\phi|-C_\mu C^\mu|\phi|+\frac{m^2}{2}\partial_{|\phi|} V&=&0~,\nonumber\\
     \partial_\mu\left(|\phi|^2C^\mu\right)&=&0~. \label{GaugeInvarEOM}
\end{eqnarray}
The third equation above is equivalent to the first equation, which can be found by applying the divergence operator to both sides. These expressions are now free from the gauge choice since both \(C^\mu\) and \(|\phi|\) do not transform under \(U(1)\)-gauge transformation. Furthermore, all of the physical fields can be straightforwardly expressed in terms of this new set of fields.

Let us introduce perturbation on the scalar field as \(\phi\rightarrow\tilde{\phi}\) with \(\tilde{\phi}=\left(f(r)+\varepsilon u(t,r)+i\varepsilon v(t,r)\right)e^{i\left(n\theta+\omega t\right)}\) where the unperturbed field is \(\phi=f(r)e^{i\left(n\theta+\omega t\right)}\), such that the unperturbed \(\arg[\phi]\) is \(n\theta+\omega t\). Similarly, we also perturb \(C_\mu\) in a straightforward manner, \(C_\mu\rightarrow\tilde{C}_\mu\), namely \(\tilde{C}_\mu=C_\mu+c_\mu\). Substituting this perturbed fields to the field equations \eqref{eom1} and \eqref{eom2}, up to first order in \(\varepsilon\), gives us three independent equations for \(c_\mu\), \(u\) and \(v\) as follows
\begin{eqnarray}\label{CperturbedEOM}
    w\left(\partial_\sigma\partial^\sigma c^\mu-\partial^\mu\partial_\rho c^\rho\right)
     -k\epsilon^{\mu\rho\sigma}\partial_\rho c_\sigma+2f^2c^\mu+4fuC^\mu+2\left(f\partial^\mu v-v\partial^\mu f\right)&=&0~,\\
     \partial_\sigma\partial^\sigma u-C^\sigma C_\sigma u-2fC_\sigma c^\sigma-2C^\sigma\partial_\sigma v-v\partial_\sigma C^\sigma+2m^2f^2V''(f^2)u&=&0~,\\
     \partial_\sigma \partial^\sigma v - C_\sigma C^\sigma v + 2C^\sigma \partial_\sigma u + u\partial_\sigma C^\sigma + 2c^\sigma \partial_\sigma f + f\partial_\sigma c^\sigma + m^2 V'(f^2)v&=&0~,\label{vperturbedEOM}
\end{eqnarray}
while the zeroth order is exactly \eqref{GaugeInvarEOM} with \(|\phi|=f(r)\). Here, the \(V'(f^2)\) notation implies the first derivative with respect to \(f^2\), \(\partial_{f^2}V(f^2)\). This set of equations are the most general gauge invariant field equations we can achieve from the Maxwell-Chern-Simons theory with axial symmetry. From this point, we further assume a potential with  vacuum at \(f_\infty\), where \(f_\infty>0\), such that \(\partial_fV(f_\infty)=0\) and \(\partial_f^2V(f_\infty)>0\). An example for this type of potential is the well-known \(|\phi|^4\) potential, \(V(|\phi|^2)=(1-|\phi|^2)^2\), which has two stationary points, \(|\phi|=1\) being the vacuum. Typical vortex solutions have a radial profile that interpolates between the two stationary points and approaches the vacuum value at large asymptotic boundary, namely \(f(r\rightarrow\infty)=f_\infty\). This asymptotic profile restricts the behaviour of \(C_\mu\) at large \(r\), which can be deduced from the charge conservation equation, namely the third equation of \eqref{GaugeInvarEOM}. Upon substitution of the perturbed fields into the charge conservation equation, we arrive at the following identity at \(r\rightarrow\infty\)
\begin{eqnarray}
    \partial_\mu(\tilde{C}^\mu|\tilde{\phi}|^2)&=&\partial_\mu(\left[C^\mu|+\varepsilon c^\mu\right]\left[\left(f+\varepsilon u\right)^2+\varepsilon^2v^2\right])\nonumber\\
    &=&\partial_\mu\left(C^\mu f^2\right)+\varepsilon\partial_\mu\left(2fuC^\mu+f^2c^\mu\right)+\dots\nonumber\\
    &\simeq&f^2_\infty\partial_\mu C^\mu+\varepsilon f_\infty\partial_\mu\left(2uC^\mu+f_\infty c^\mu\right)+\dots .
\end{eqnarray}
Isolating only the zeroth-order and first-order terms, we arrive at \(\partial_\mu C^\mu=0\) from the zeroth-order terms and \(\partial_\mu\left(2uC^\mu+f_\infty c^\mu\right) =0\) from the first-order terms. Furthermore, we also employ the boundary conditions of \(A_\mu\) to construct the boundary conditions of \(C^\mu\). From the ansatz, we found that \(C_0=A_0+\partial_t\arg [\phi]=g+\omega\), \(C_r=A_r+\partial_r\arg [\phi]=0\), and \(C_\theta=A_\theta+(\partial_\theta\arg [\phi]/r)=(2n-h)/r\), which implies that \(\lim_{r\rightarrow\infty}C_\mu=0\) if \(A_\mu\) satisfies the usual boundary conditions for a vortex, given in eq. \eqref{ABoundCond}. To sum up, we have \( C^\mu\simeq0,\) and \(\partial_\mu c^\mu \simeq0\) at large \(r\). From here, we write \(f_\infty=1\) without any loss of generality. Evaluating equations (\ref{CperturbedEOM}-\ref{vperturbedEOM}) at large \(r\) using the identities for \(C_\mu\) and \(c_\mu\) above, gives the field equations at large \(r\),
\begin{eqnarray}
    w\partial_\sigma\partial^\sigma c^\mu
     -k\epsilon^{\mu\rho\sigma}\partial_\rho c_\sigma+2c^\mu+2\partial^\mu v&\simeq&0~,\\
     \partial_\sigma\partial^\sigma u+2m^2V''(f^2_\infty)u&\simeq&0~,\\
     \partial_\sigma \partial^\sigma v &\simeq&0~.
\end{eqnarray}
We now recast the equations in polar coordinates and impose axial symmetry such that \(u,\) \(v,\) and all components of \(c_\mu\) only depend on \(t\) and \(r\). Since we evaluate the equations above at large \(r\), all terms that spatially decay as either \(r^{-1}\) or faster can be omitted. This approach gives us the following set of linear equations at large \(r\),

\begin{align}
 \partial_t^2c_0-\partial_r^2c_0
+\frac{k}{w}~\partial_rc_\theta+\frac{2}{w}\left(\partial_t v
+c_0\right)&\simeq
0,
\\[0.8em]
\partial_t^2c_r-\partial_r^2c_r+\frac{k}{w}\,\partial_t c_\theta-\frac{2}{w}\left(\partial_r v- c_r\right)
&\simeq0,
\\[0.8em]
\partial_t^2c_\theta-\partial_r^2c_\theta-\frac{k}{w}(\partial_t a_r+\partial_r a_0)+\frac{2}{w} a_\theta
&\simeq0,\\[0.8em]
\partial_t^2 u
-\partial_r^2 u
+2m^2 V''(1)
u&\simeq0,
\\[0.8em]
\partial_t^2 v
-\partial_r^2 v&\simeq0,
\end{align}
that is the same large \(r\) approximation from \((u,v,a_\mu)\) that is used to derive the dispersion relation in the main paper. Our demonstration revealed that, although the field equations for \(C_\mu\) and \(A_\mu\) are indeed different at finite \(r\), the large \(r\) approximation for \(A_\mu\) equipped with Lorenz gauge turns out to be exactly the same as the large \(r\) approximation for \(C_\mu\), provided that both satisfy the boundary conditions of an axially symmetric vortex. Thus, the dispersion relation and the exponentially growing modes found for \(k>2\sqrt{2w}\) are physical since they can be deduced using \(C_\mu\) that is free from gauge choices. However, this correspondence between the asymptotic behaviour of \(A_\mu\) and \(C_\mu\) is not guaranteed in a more general setup beyond an axially symmetric single-vortex background. The calculation using gauge-invariant \(C_\mu\) is more reliable for formulating the stability conditions in the more general setups, since it does not rely on the gauge choice.

\section{Convergence to continuous spectrum}\label{convergence}
\begin{figure}
    \centering
    \includegraphics[width=0.85\linewidth]{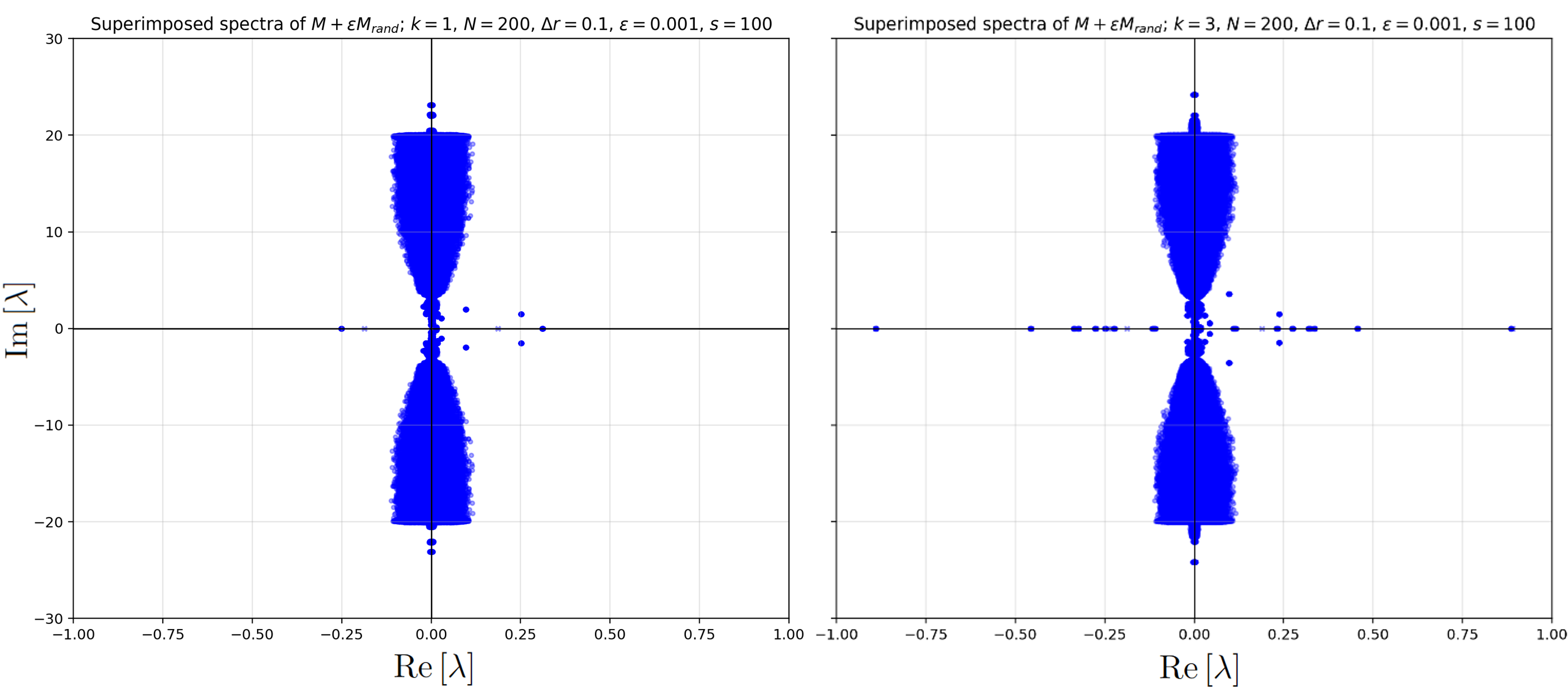}
    \includegraphics[width=0.85\linewidth]{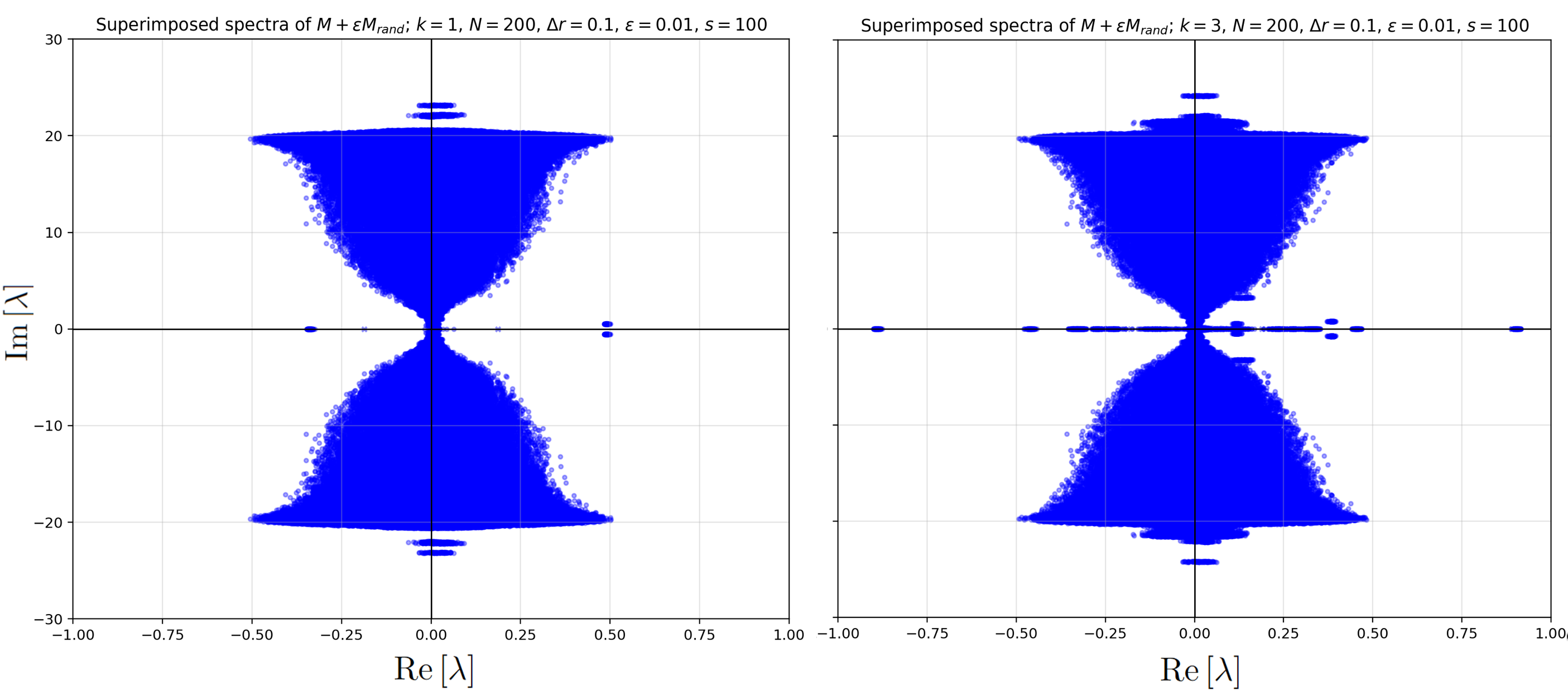}
    \caption{Pseudo-spectra generated by perturbing matrix \eqref{finitersystem} with 100 randomly generated matrix of the same dimension. The unperturbed spectrum is more sensitive to perturbation for eigenvalues with large imaginary part.}
    \label{fig:pseudospectra}
\end{figure}
The discretized linear operator associated with the present eigenvalue problem exhibits a broad $\varepsilon$-pseudospectrum even for very small values of $\varepsilon$, indicating strong nonnormality and pronounced spectral sensitivity \cite{TrefethenEmbree,Davies2007}. This behavior is likely connected to the presence of first-order spatial derivative terms in the linearized operator. Such transport-like contributions are well known to generate non-normality, substantial resolvent growth, and enhanced sensitivity to perturbations, particularly for oscillatory or spatially extended modes \cite{TrefethenEmbree,Schmid2007}. Consequently, small perturbations arising from spatial discretization, domain truncation, round-off errors, or weak nonlinear interactions may induce substantial shifts in the computed eigenvalues, including the appearance of isolated spectral branches that are not robust under mesh refinement or changes in numerical resolution.

In the present computations, the spectral components located near the imaginary axis and farther from the origin exhibit the strongest sensitivity under discretization and domain refinement, suggesting that these branches are dominated primarily by pseudospectral effects associated with nonnormal transport dynamics. By contrast, the isolated eigenvalues localized near the real axis remain comparatively robust, indicating that they likely correspond to genuine localized point-spectrum modes of the underlying continuum operator. This distinction is consistent with the general behavior of nonnormal differential operators, for which extended high-frequency modes are considerably more sensitive to perturbations than spatially localized states \cite{Davies2007,TrefethenEmbree}.

It is expected that the discretization of the matrix \eqref{finitersystem} for numerical procedures generates pseudo-spectra that actually have no physical relevance to the original system. We can simulate how the pseudo-spectra could appear during the numerical procedure by calculating the spectrum of eigenvalues of the perturbed matrix. Suppose that the matrix in eq. \eqref{finitersystem} is denoted by \(M\), the perturbed matrix is given bt \(M+\varepsilon M_{\text{rand}}\) where \(\varepsilon\) is small parameter and \(M_{\text{rand}}\) is a random matrix with the same dimension as \(M\). The entries of \(M_{\text{rand}}\) take values in the interval \([-1,1]\), and this random matrix represents errors that could develop when we apply the numerical procedure. We plotted the spectra of the eigenvalues of \(M+\varepsilon M_{\text{rand}}\) for several randomly generated \(M_{\text{rand}}\) and found that, the matrix \(M\) (discretized by procedure described in section \ref{finiterspectrum}) is quite sensitive to the perturbation due to the terms containing first derivatives in \(r\). As demonstrated in Fig. \ref{fig:pseudospectra}, the pseudo-spectra generated from perturbed \(M\) are distributed widely, away from the unperturbed spectrum in Fig. \ref{fig:FiniteRSpectrum}. According to this simulation, the spectrum is less sensitive near the real \(\lambda\) axes, and the sensitivity grows as the imaginary part of \(\lambda\) grows. This implies that we cannot be sure about the validity of the numerically calculated \(\lambda\) of \(M\) for \(\lambda\) with a larger imaginary part, since the possibility that it belongs to the pseudo-spectra is higher. On the other hand, this simulation shows that we can be confident in the completely real eigenvalues that are also predicted by the continuous spectrum when \(k>2\sqrt{2w}\). Hence, these real eigenvalues that generate instabilities are physically relevant. The full time-evolution discussed in section \ref{VortexTimeEvolution} is in agreement with this observation because the vortex for \(k<2\sqrt{2w}\) is indeed stable, implying that \(\lambda\) with small real part in the numerically calculated spectrum of \(M\) are not physically relevant. 

The approximate spectrum for finite \(r\) numerically computed in section \ref{finiterspectrum} does converge to the continuous spectrum found using the asymptotic limit in section \ref{largerspectrum}. The convergence happens when the maximum \(r=r_\text{max}\), at which the asymptotic boundary conditions are applied, is taken to be large.

\begin{figure}
    \centering
    \includegraphics[width=0.5\linewidth]{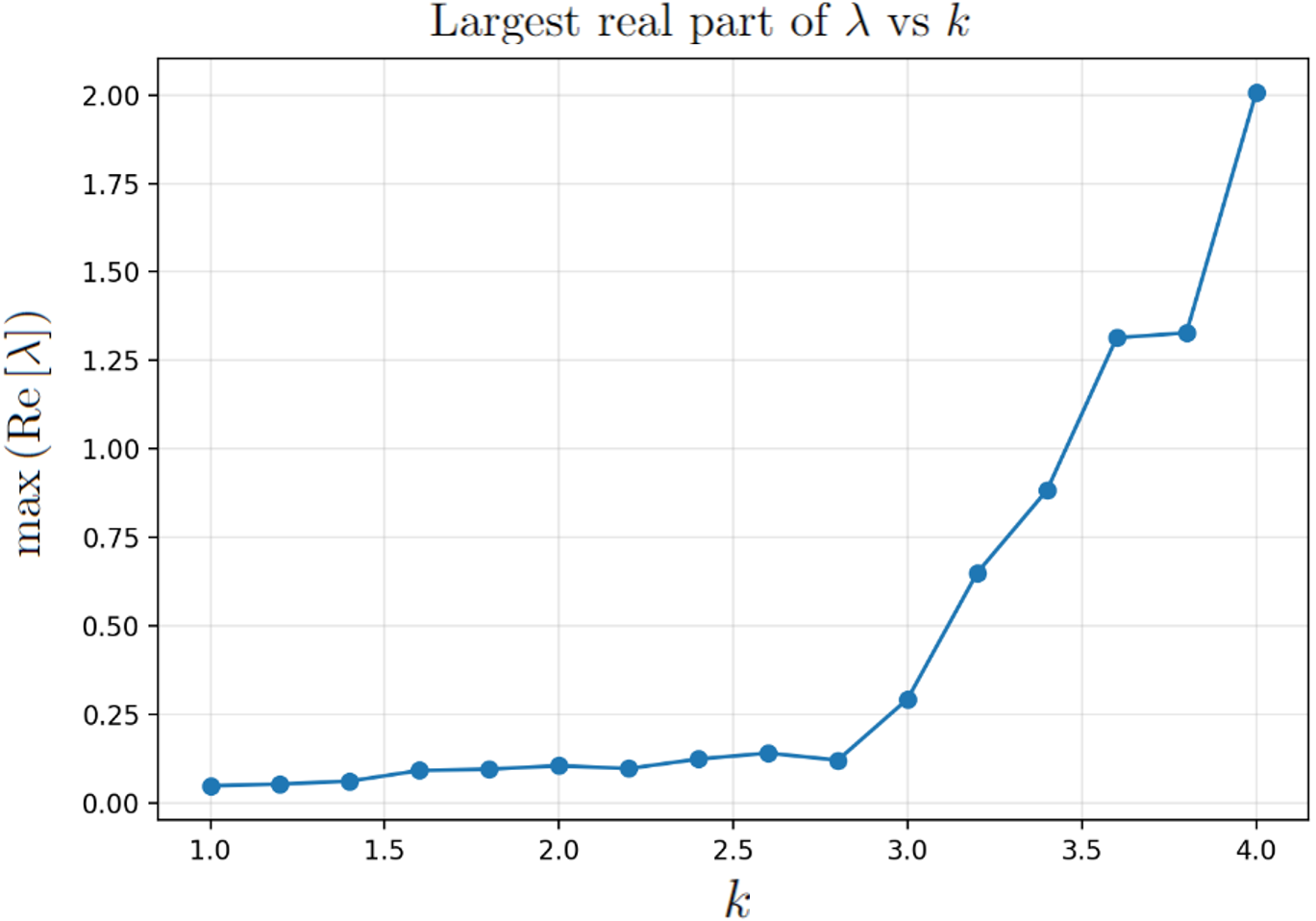}
    \caption{The largest numerical values of real part of eigenvalues plotted against \(k\), for \(w=1\) cases. Significant instability arises at \(k\approx2\sqrt{2}\), as predicted by the asymptotic spectrum with \(w=1\).}
    \label{fig:HighRealLambda}
\end{figure}

We further confirm the spectrum convergence to the continuous spectrum by plotting the highest real part of \(\lambda\) in Fig. \ref{fig:HighRealLambda} under variations of \(k\) with \(w=1\) kept constant. In the continuous spectrum, the highest real part is zero for \(k\) in \([0,2\sqrt{2w}]\), and it grows as \(\sqrt{(k^2-8w)/(8w^2)}\) when \(k\) enters the unstable regime, namely \((2\sqrt{2w},\infty)\). For sufficiently large \(k\), the highest real part of \(\lambda\) grows almost linearly with \(k\). A similar behaviour is also observed in Fig. \ref{fig:HighRealLambda}, where the highest \(\text{Re}[\lambda]\) stays close to zero up to \(k=2.8\) and starts to grow almost linearly starting from \(k=3\). Under variation of the discretization parameters, the small real parts for \(k\) in the stable regime shrink as \(\Delta r\) decreases or \(r_{\text{max}}\) increases.

\section{Numerical method for time evolution}\label{timeIntegrateMethod}
In this section of the appendix, we describe the computational procedure to simulate the time evolution of the solution in both stable and unstable regimes.
To solve the full dynamical system (\ref{FUllDynamicsA0}-\ref{fullDynamicsPhi}), we first recast the system into first-order in \(t\) by writing the scalar field as
\begin{equation}
    \phi\equiv F(t,r,\theta)e^{i\Gamma(t,\theta)}
\end{equation}
and defining a momentum-like quantity,
\begin{equation}
    \Pi_0\equiv\partial_tA_0,~\Pi_r\equiv\partial_tA_r,~\Pi_\theta\equiv\partial_tA_\theta,~\Pi_F\equiv\partial_tF,~\Pi_\Gamma\equiv\partial_t\Gamma.
\end{equation}
In this notation, (\ref{FUllDynamicsA0}-\ref{fullDynamicsPhi}) becomes
\begin{align}\label{DerivativePi0inT}
\partial_t\Pi_0&=\Delta A_0
- \frac{k}{w}\left(\frac{1}{r}\partial_r(rA_\theta)-\frac{1}{r}\partial_\theta A_r\right)
+ \frac{2}{w}F^2(\Pi_\Gamma-A_0), \\[1ex]
\partial_t \Pi_r&=\Delta A_r-\frac{A_r}{r^2}-\frac{2}{r^2}\partial_\theta A_\theta
- \frac{k}{w}\left(\Pi_\theta+\frac{1}{r}\partial_\theta A_0\right)
- \frac{2}{w}F^2(\partial_r\Gamma+A_r), \\[1ex]
\partial_t\Pi_\theta&=\Delta A_\theta-\frac{A_\theta}{r^2}+\frac{2}{r^2}\partial_\theta A_r
+ \frac{k}{w}\left(\Pi_r+\partial_r A_0\right)
- \frac{2}{w}F^2\left(\frac{1}{r}\partial_\theta\Gamma+A_\theta\right), \\[1ex]
\partial_{t}\Pi_F&=\Delta F
+F\left[\left(\Pi_\Gamma+A_0\right)^2
-\left(\partial_r\Gamma-A_r\right)^2
-\left(\frac{\partial_\theta\Gamma}{r}-A_\theta\right)^2\right]
-m^2 F\frac{d V}{d(F^2)}
 ,\\[1ex]
\partial_{t}\Pi_\Gamma&=\Delta\Gamma
-2\frac{\Pi_F}{F}\left(\Pi_\Gamma+A_0\right)
+2\frac{F_r}{F}\left(\Gamma_r-A_r\right)
+\frac{2F_\theta}{rF}
\left(\frac{1}{r}\Gamma_\theta-A_\theta\right)
 .\label{DerivativePiGammainT}
\end{align}
Let \(\Pi\) be a vector with components \(\Pi=(\Pi_0,\Pi_r,\Pi_\theta,\Pi_F,\Pi_\Gamma)\).
Using forward finite difference for the time derivative with temporal slice \(\delta t\), we can integrate \(\Pi\)s numerically to get the time evolutions of the fields,
\begin{equation}
    \partial_t\begin{bmatrix}
        A_0\\A_r\\A_\theta\\F\\\Gamma
    \end{bmatrix}=~\Pi (t,r,\theta ),
\end{equation}
which should be solved simultaneously with \(\Pi(t,r,\theta)\) from \(\partial_t\Pi\) in eq. (\ref{DerivativePi0inT}-\ref{DerivativePiGammainT}) above.
The initial condition is chosen to be \(\Pi(0,r,\theta)=0\) which correspond to the stationary solutions as the initial data for the fields, i. e.
\begin{equation}
    A_r (0,r,\theta)=0,~ A_\theta (0,r,\theta) = -\frac{1}{r}\big(h(r)-n\big),~A_0(0,r,\theta) = g(r),~\phi(0,r,\theta)=f(r)e^{in\theta}.\end{equation}
However, in order to simplify the computational procedure, we estimate \(f,~g\) and \(h\) from their asymptotic solutions. These asymptotic solutions satisfy approximations (\ref{AsymptoticSolg}-\ref{AsymptoticSolf}). The use of the asymptotic solutions as the initial data also introduces small deviation from the exact stationary solutions. This small deviation allows the system to evolve, either away from the stationary solutions for unstable regime, or approaching the stationary solutions for stable regime. In this work, we restrict the scope by assuming azimuthal symmetry during the time evolution, indicated by the choice of initial data from the stationary vortex solution, described above. As such, the more general cases where azimuthal symmetry can be broken during the time evolution are beyond the scope of this work and require a more careful treatment since there are extra constraints from the periodicity of the fields at \(\theta=0\) and \(\theta=2\pi\). These general cases will be considered in future works. 

\bibliography{ref.bib}



\end{document}